# ТЕПЛОВЫЕ ВАКАНСИИ В 1D ЦЕПОЧКАХ АДСОРБАТА КСЕНОНА В КАНАВКАХ СВЯЗОК НАНОТРУБОК

**М. И. Багацкий, М. С. Барабашко, В. В. Сумароков**
*Физико-технический институт низких температур им. Б. И. Веркина НАН, 61103 Харьков, Украина*
barabaschko@ilt.kharkov.ua

Измерена теплоемкость $C_P$ квази-одномерных цепочек адсорбата Xe в канавках на внешней поверхности связок закрытых одностенных углеродных нанотрубок в интервале температур 2 - 55 К. Экспериментальные данные сопоставили с теоретическими значениями $C_V$. Резкое увеличение разности $C_P$ - $C_V$ выше 30 К объясняется в рамках модели образования одиночных тепловых вакансий в 1D цепочках адсорбата Xe. Определены энтальпия, энтропия и концентрация тепловых вакансий.

Уникальная структура связок одностенных углеродных нанотрубок закрытых на концах (з-ОУНТ) позволяет получать низкоразмерные системы (квази- 1D, -2D, -3D), образованные адсорбатами. Физические свойства таких систем интенсивно исследуются теоретически и экспериментально [1-11].

В предыдущей нашей работе [1] сообщаются результаты экспериментального исследования теплоемкости $C_P$ плотных 1D цепочек адсорбатов ксенона в канавках связок з-ОУНТ ниже 30 К. Эксперимент совпадает с теоретическим расчетом фононной теплоемкости $C_V$ [12] ниже 8 К. А выше 8 К появляется, увеличивающееся с температурой, расхождение между ними.

В теоретической работе [13] показано, что упорядоченная бесконечно длинная 1D фаза является не устойчивой. Процессы образования, упорядочения и распада 1D цепочек ограниченной длины с количеством атомов $N \le 1000$ были исследованы методом компьютерного моделирования в работе [14]. Влияние внешнего потенциального поля, в котором находятся цепочки, в теории не учитывали. Химический потенциал конечных упорядоченных цепочек является функцией длины цепочек и температуры. Установлено, что при низких температурах упорядоченные 1D цепочки ограниченной длины могут существовать. С повышением температуры интенсивность тепловых возбуждений увеличивается, и в цепочке возрастают флуктуации плотности атомов. С ростом флуктуаций плотности атомов, вначале, происходит разупорядочение атомов на концах цепочек а, затем, - распад длинных цепочек на более короткие цепочки (фрагментация цепочек). Температура начала фрагментации $T_0$ повышается с уменьшением длины цепочек [14].

Целью настоящей работы является исследование калориметрическим методом влияния образования вакансий в 1D цепочках адсорбата ксенона в канавках на внешней поверхности связок з-ОУНТ на теплоемкость 1D цепочек.

Теплоемкость калориметра и связок з-ОУНТ, с физически адсорбированными во внешние канавки 1D цепочками атомов Xe ($C_{ad+Xe}$), была измерена на адиабатическом калориметре [15] в области температур от 2 до 55 К. Цилиндрический образец связок з-ОУНТ (высотой 7.2 мм, 10 мм в диаметре) был приготовлен сжатием пластин под давлением 1.1 ГПа. Пластины (~0.4 мм) получали прессованием порошка ('Cheap Tubes') з-ОУНТ под давлением 1.1 ГПа [16]. Порошок получен методом CVD. Он содержит более 90 % вес. связок з-ОУНТ, а также и другие аллотропные формы углерода (фуллерит, многостенные нанотрубки, аморфный углерод), и около 2.9 % вес. катализатора Co. Средний диаметр нанотрубок равен 1.1 нм; средняя длина связок равна 15 мкм; среднее количество нанотрубок в связке, согласно оценке [17], составляло 127. Масса образца равна 716.00 ± 0.05 мг.

Тщательная очистка образца от газовых примесей и измерения теплоемкости $C_{ad}$ калориметра с чистым образцом ("addenda") были проведены перед насыщением связок з-ОУНТ ксеноном [17]. В рамках геометрической модели оценили массу Xe, необходимую для плотного заполнения всех канавок в связках

(без структурных вакансий). Количество ксенона 3.19·10⁻⁴±5·10⁻⁶ моль, использованное в эксперименте, было определено PVT- методом. Химическая чистота ксенона была 99.98% (0.01% $N_2$, 0.01% Kr). Отношение $N_{Xe}/N_C$ = 0.55%, где $N_{Xe}$, $N_C$ – количество атомов ксенона и углерода в образце, соответственно. Другие экспериментальные детали описаны в [1,15,17].

Теплоемкость при постоянном давлении $C_P$ физически адсорбированных 1D цепочек атомов Xe во внешние канавки связок з-ОУНТ, была определена вычитанием теплоемкости $C_{ad}$ из общей теплоемкости $C_{ad+Xe}$. Отношение $C_P/C_{ad}$ ≈ 160 % при 2.6 K, ≈ 20% при 30 K и ≈ 12% при 55 K. Случайная ошибка в определении значений $C_P$ не превышает ± 20% при 2.2 K, ± 5% в температурном интервале 10 – 30 K и увеличивается до ±8% при 55 K. Основной вклад в систематическую погрешность вносит неопределенность в количестве связок з-ОУНТ.

Температурные зависимости экспериментальной $C_P(T)/R$ и теоретической $C_V(T)/R$ теплоемкостей, нормированных на газовую постоянную $R$, показаны на рисунке 1. Фононная теплоемкость при постоянном объеме $C_V(T)$ рассчитана до 40 K в работе [12]. Ниже 8 K экспериментальная и теоретическая кривые совпадают. Ниже 4 K наблюдается линейный участок $C_P(T)/R$, обусловленный продольной модой. Выше этой температуры становятся заметными вклады поперечных мод. Выше 8 K наблюдается отклонение $C_P(T)/R$ от $C_V(T)/R$. Высказано предположение, что в области 8 – 30 K разность $\Delta C(T)/R = (C_P - C_V)/R$ обусловлена, в основном, вкладом теплового расширения 1D цепочек $C_\alpha(T)$ [1]. При 40 K $C_V(T)/R$ близка к классическому пределу $C_V(T)/R = 3$. Обращает на себя внимание тот факт, что на кривая $C_P(T)/R$ вблизи 30 K имеется излом и выше 30 K резко возрастает.

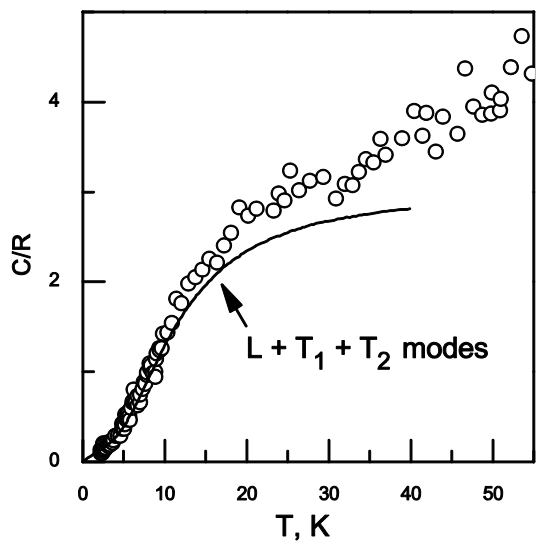

Рис. 1. Температурная зависимость теплоемкости 1D цепочек атомов Xe, адсорбированных во внешних канавках связок c-SWNT. Эксперимент ($C_P(T)/R$): открытые кружки. Теория ($C_V(T)/R$) [12]: сплошная кривая для продольной и поперечных L+T1+T2 мод.

Кривая температурной зависимости $\Delta C(T)/R$ между экспериментальными сглаженными $C_P$ и теоретическими $C_V$ величинами теплоёмкостями 1D цепочек атомов Xe показана на рис. 2. Видно, что поведение $\Delta C(T)$ при температурах ниже и выше 30 K качественно отличается.

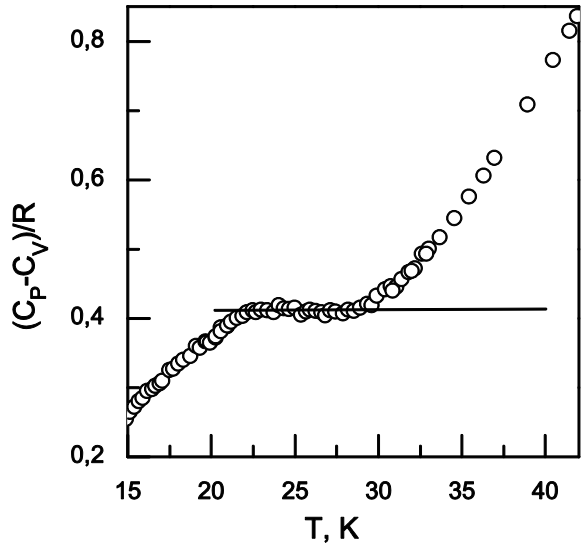

Рис. 2. Температурная зависимость разности $\Delta C(T)/R = (C_P - C_V)/R$ между теоретической $C_V/R$ и сглаженной экспериментальной $C_P/R$ кривыми теплоёмкости 1D цепочек атомов Xe. Экстраполяция до 40 K линейной зависимости $C_\alpha(T)$, наблюдаемой в интервале температур 22- 28 K, показана сплошной линией.

Мы предположили, что $\Delta C(T)$ выше 30 K обусловлено вкладами как теплового расширения 1D цепочек $C_\alpha(T)$, так и тепловых вакансий $C_{vac}(T)$. Из-за тепловых флуктуаций возможна локальная конфигурационная перестройка: атомы Xe из 1D цепочек могут перейти в ближайшие свободные позиции двух вторичных цепочек трехцепочечной структуры в канавках (см. рис.3). При этом, в нижней цепочке образуется вакансия (свободный узел), изменяются расстояния в цепочке между ближними к вакансии атомами Xe, происходит локальная деформация цепочки вблизи вакансии. Образование вакансий приводит к распаду длинных цепочек на более короткие (фрагментации цепочек). Атомы Xe во вторичных цепочках находятся в потенциальных ямах, сформированных атомами углерода нанотрубки и ближайшими атомами Xe вблизи вакансии. Сильный эффект (максимум вблизи 60 K), связанный, по мнению авторов [18], с конфигурационной перестройкой атомов ксенона на внешней поверхности связок с $N_{Xe}/N_C = 1.64\,\%$, наблюдали в радиальном тепловом расширении.

На рис. 2 видно, что от 22 K до 28 K наблюдается слабая линейная зависимость $C_\alpha(T)$. Мы предположили, что и выше 30 K сохраняется та же зависимость $C_\alpha(T)$. Экстраполяция до 40 K линейной зависимости $C_\alpha(T)$ показана на рис. 2 сплошной линией. Таким образом, выше 30 K вклад тепловых вакансий $C_{vac}(T)$ в температурную зависимость $\Delta C(T)$ равен $C_{vac}(T) = \Delta C(T) - C_\alpha(T)$.

Фазы адсорбатов во внешних канавках связок з-ОУНТ и на поверхности были рассмотрены теоретически в работах [19-24]. В работах [23,24] показано, что $U_{1G} > U_{2G} > U_{OS}$, где $U_{1G}$, $U_{2G}$, $U_{OS}$ – энергии связи атомов Xe на дне канавки, во вторичных цепочках и на поверхности нанотрубок, соответственно. Образование атомами Xe трехцепочечной структуры во внешних канавках связок з-ОУНТ было подтверждено при измерениях изотерм адсорбции ксенона [25,26]. В работах Т. Н. Анцыгиной и др. [8,10,20] теоретически изучена низкотемпературная термодинамика адсорбатов во внешних канавках связок з-ОУНТ. Фононная теплоемкость при постоянном объеме $C_V$ трехцепочечной фазы, образованной адсорбированными молекулами $CH_4$ в канавках связок з-ОУНТ, рассчитана в [27].

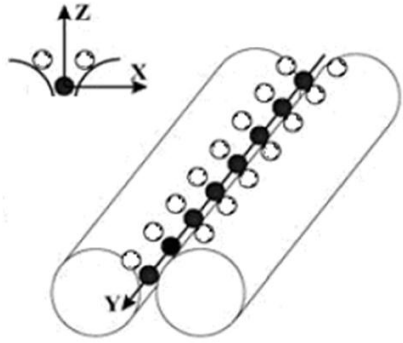

Рис. 3. Схема трехцепочечной структуры адсорбата во внешней канавке связки з-ОУНТ между двумя нанотрубками. 1D цепочка атомов Xe на дне канавки (черные кружки) и свободные позиции вторичных цепочек (светлые кружки), в которые возможен переход атомов Xe при повышении температуры.

В случае плотной цепочки в канавке средней длины $L = 15$ мкм атомы Xe заполняют все узлы, количество которых определяется выражением $N_s = L/a$, где $a$ – постоянная 1D цепочки. Предположив, что $a = 4.336$ Å (расстояние между ближайшими соседними атомами в ГЦК решетке кристаллического Xe при $T=0$ К [28]) оценили количество атомов Xe в плотной цепочке $\approx 3.5 \cdot 10^4$. При малых концентрациях вакансий ($n \ll 1$):

$$n(T) = N_{vac}/(N_{vac} + N_{Xe}) \approx N_{vac}/N_{Xe}, \tag{1}$$

где $N_{vac}$ - количество тепловых вакансий в цепочке. В случае, когда взаимодействием между вакансиями можно пренебречь, энтальпия $h$ и энтропия $s$ образования одной вакансии не зависят от температуры. Термодинамический потенциал $g$ образования одной вакансии равен $g \equiv h - Ts$ [28]. Для определения $h$, $s$, $n(T)$ и $C_{vac}(T)$ использовали выражения, применяемые в случае отвердевших инертных газов [28]:

$$n \equiv e^{-g/kT} \approx \frac{N_{vac}}{N_{Xe}}, \tag{2}$$

$$C_{vac} = N_{vac}\frac{h^2}{kT^2} = N_{Xe}\frac{h^2}{kT^2}e^{-g/kT} = N_{Xe}\frac{h^2}{kT^2}e^{-(h-Ts)/kT}, \tag{3}$$

где $k$ – постоянная Больцмана.

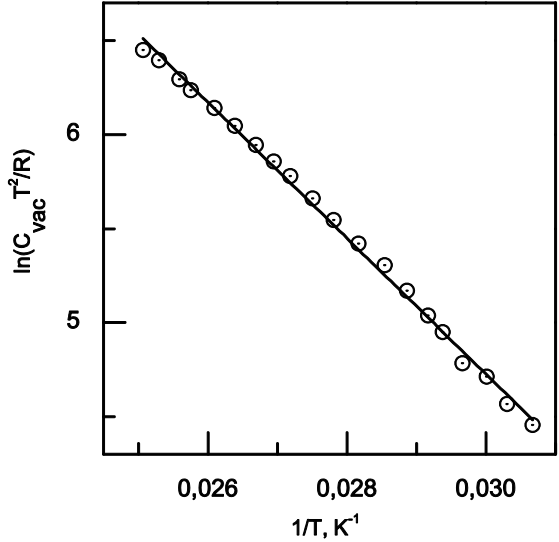

Рис. 4. Зависимость $\ln(\Delta C_{vac} \cdot T^2/R)$ от $1/T$. Эксперимент - черные квадраты. Прямая линия – аппроксимация.

После логарифмирования и преобразования выражения (3) получим

$$\ln(C_{vac} \cdot T^2/R) = -h/kT + s/k + 2\ln(h/k). \qquad (4)$$

График зависимости $\ln(C_{vac} \cdot T^2/R)$ от обратной температуры приведен на рис. 4. Отметим, что в интервале температур 33 – 40 К зависимость $\ln(C_{vac} \cdot T^2/R)$ от $1/T$ близка к линейной. Из угла наклона прямой, приведенной на рис. 4, определили энтальпию образования вакансии $h/k \cong 361$ К ($H \cong 3000$ Дж/моль). Энтропия ($s/k$) образования одной вакансии не зависит от температуры: $s/k = 3.75 \pm 0.05$. Концентрация вакансий при температурах 33 К и 40 К равна $8 \cdot 10^{-4}$ и $5 \cdot 10^{-3}$, соответственно. Полученные термодинамические величины образования одиночных вакансий представляются реалистичными. Значение $H$ близко к величине разности энергии связи атома Xe в цепочке на дне канавки $U_{1G}$ и на поверхности связок $U_{1G} - U_{OS} = 2.01$ кДж/моль [23]; $U_{1G} - U_{OS} = 1.5$ кДж/моль [24]. Отметим, что экспериментальное определение $U_{1G}$ связано со значительными трудностями. Энтальпия образования вакансии $h/k$ примерно на порядок меньше, чем энергия связи $\approx 3300$ К [23] атома Xe с углеродными нанотрубками в канавке, и в 3-4 раза меньше, чем энтальпия образования вакансии в твердом Xe (1100 К<$h/k$<1250 К) [29]. В твёрдом Xe каждый атом имеет 12 ближайших соседних атомов, а в 1D цепочках – 2 атома. Поэтому $h/k$ в твердом Xe в несколько раз больше, чем в цепочке. Равновесное давление газа Xe в объеме над связками з-ОУНТ при температуре 55 К меньше $1 \cdot 10^{-6}$ мм рт.ст. Поэтому влиянием сублимации атомов на теплоемкость плотных 1D цепочек атомов Xe можно пренебречь.

Температура ~ 28 К, при которой вклад процессов распада плотной цепочки на более короткие цепочки, вследствие образования тепловых вакансий, становится заметным в теплоемкости $\Delta C(T)$ (см. рис.2), близка к величине $T_0$ [14], определяемой из уравнения:

$$\kappa T_0 = \frac{\varepsilon}{\ln N}, \qquad (5)$$

где ε – глубина потенциальной ямы в потенциале Леннарда-Джонса, $N$ – количество атомов в цепочке при $T < T_0$, $T_0$ – температура при которой процессы распада цепочек на более короткие начинают вносить заметный вклад в теплоемкость. Согласно [14] плотные цепочки конечной длины, образованная $N$ атомами, могут оставаться упорядоченными при низких температурах $T<T_0$. В случае нашего образца связок з-ОУНТ в канавке средней длины $L$= 15 мкм находится $N \approx 3.5*10^4$ атомов Хе. Глубина потенциальной ямы в потенциале Ленард-Джонса для ксенона равна 230.4 К [30]. Воспользовавшись уравнением (5), получили $T_0 \approx 22$ К. Отметим, что в уравнении (5) не учитываются влияния на $T_0$ тепловых возбуждений в цепочке и потенциального поля, в котором находится цепочка.

Предложенная нами модель механизма разрушения плотных 1D цепочек атомов Хе на дне канавок на внешней поверхности связок з-ОУНТ, вследствие образования тепловых вакансий, представляется реалистичной.

С повышением температуры выше 40 К увеличивается концентрация тепловых вакансий, и необходимо учитывать взаимодействие между вакансиями. При более высоких температурах атомы Хе могут переходить из вторичных цепочек в канавках на внешнюю поверхность связок з-ОУНТ.

В заключение отметим, что низкотемпературная динамика цепочек адсорбата Хе во внешних канавках связок закрытых одностенных углеродных нанотрубок такая же, как в случае 1D кристалла. Экспериментальная $C_P(T)$ и теоретическая $C_V(T)$ кривые хорошо согласуются ниже 8 К. Резкое увеличение температурной зависимости разности $C_P - C_V$ выше 30 К, по нашему мнению, обусловлено вкладом тепловых вакансий, которые образуются в плотных 1D цепочках Хе. Полученные значения термодинамических величин энтальпии $h$ и энтропии $s$ образования одной вакансии, а также концентрации вакансий представляются реалистичными.

Список литературы:

[1] M. I. Bagatskii, V. G. Manzhelii, V. V. Sumarokov, and M. S. Barabashko, Fiz. Nizk. Temp. **39**, 801 (2013). [Low Temp Phys. **39**, 618 (2013)].

[2] M. I. Bagatskii, M.S . Barabashko, V. V. Sumarokov, Fiz. Nizk. Temp. **39**, 568 (2013). [Low Temp. Phys. **39**, 441 (2013)].

[3] P. Kowalczyk, P. A. Gauden, and A. P. Terzyk, J. Phys. Chem. B **112**, 8275 (2008).

[4] G. Garberoglio, and J. K. Johnson, ACS Nano **4**, 1703 (2010).

[5] Y. H. Kahng, R. B. Hallock, and E. Dujardin, Phys. Rev. B **83**, 115434 (2011).

[6] E. V. Manzhelii, I. A. Gospodarev, S. B. Feodosyev, and N. V. Godovanaja, in 9th Int. Conf. on Cryocrystals and Quantum Crystals (CC2012), Odessa, 63 (2012).

[7] М. В. Харламова, УФН. **183**, 1146 (2013).

[8] T. N. Antsygina, I. I. Poltavsky, K. A. Chishko, T. A. Wilson, and O. E. Vilches, Fiz. Nizk. Temp. **31**, 1328 (2005) [Low Temp. Phys. **31**, 1007 (2005)].

[9] А. В. Елецкий, УФН. **174**, 1191 (2004).

[10] T. N. Antsygina, I. I. Poltavsky, and K. A. Chishko, Phys. Rev B **74**, 205429 (2006).

[11] K. Percus, Mol. Phys. **100**, 2417 (2002).

[12] A. Šiber, Phys. Rev. B **66**, 235414 (2002).

[13] Л. Д. Ландау, ЖЭТФ **7**, 627 (1937).

[14] J. M. Phillips and J. G. Dash, J. Stat. Phys. **120**, 721 (2005).

[15] M. I. Bagatskii, V. V. Sumarokov, A. V. Dolbin, Fiz. Nizk. Temp. **37**, 535 (2011) [Low Temp Phys. **37**, 424 (2011)].

[16] A. V. Dolbin, V. B. Esel'son, V. G. Gavrilko, V. G. Manzhelii, N. A. Vinnikov, S. N. Popov, and B. Sundqvist, Fiz. Nizk. Temp. **34**, 860 (2008) [Low Temp. Phys. **34**, 678 (2008)].

[17] M. I. Bagatskii, M. S. Barabashko, A. V. Dolbin, V. V. Sumarokov, Low Temp. Phys. **38**, 523 (2012) [Fiz. Nizk. Temp. **38**, 667 (2012)].

[18] A. V. Dolbin, V. B. Esel'son, V. G. Gavrilko, V. G. Manzhelii, N. A. Vinnikov, S. N. Popov, N. I. Danilenko, and B. Sundqvist, Fiz. Nizk. Temp. **35**, 613 (2009).

[19] T. N. Antsygina, I. I. Poltavsky, and K. A. Chishko, J. Low Temp. Phys. **148**, 821 (2007).

[20] S. M. Gatica, M. J. Bojan, G. Stan, and M. W. Cole, J. Chem. Phys. **114**, 3765 (2001).

[21] S. M. Gatica, M. M. Calbi, R. D. Diehl, M. W. Cole, J Low Temp Phys **152**, 89(2008).

[22] A. D. Lueking and M. W. Cole, Phys. Rev. B **75**, 195425 (2007).

[23] A. J. Zambano, S. Talapatra, and A. D. Migone, Phys. Rev. B **64**, 075415 (2001).

[24] H. Ulbricht, J. Kriebel, G. Moos, and T. Hertel, Chem. Phys. Lett. **363**, 252 (2002).

[25] S. Talapatra, A. D. Migone, Phys. Rev. Lett. **87**, 206106 (2001).

[26] S. Talapatra, V. Krungleviciute, and A. D. Migone, Phys. Rev. Let. **89**, 246106, (2002).

[27] M. K. Kostov, M. M. Calbi, and M. W. Cole, Phys. Rev. B. **68**, 245403 (2003).

[28] *Rare Gas Solids*, M. L. Klein and J. A. Venables (eds.), Academic Press, London-New York-San Francisco, Vol. **2**., 1252 (1977).

[29] P. R. Granfors, A. T. Macrander, and R. O. Simmons, Phys. Rev. B **24**, 4753 (1981).

[30] V. G. Manzhelii, A. I. Prokhvatilov, I. Ya. Minchina, L. D. Yantsevich, *Handbook of Binary Solutions of Cryocrystals*, begell house, New York - Wallingford (UK), 236 (1996).